\documentclass[twocolumn,showpacs,preprintnumbers,amsmath,amssymb]{revtex4}

\usepackage{graphicx}
\usepackage{dcolumn}
\usepackage{bm}

\begin{document}

\preprint{APS/123-QED}


\title{Plasma effects in   a micromachined floating-gate high-electron-mobility
transistor
} 
\author{Y.~Hu and I.~Hagiwara}
\affiliation{Department of Mechanical Sciences and Engineering,
Tokyo Institute of Technology,
 Tokyo 152-8552, Japan}
\author{I.~Khmyrova, M.~Ryzhii, and V.~Ryzhii} 
\email{v-ryzhii@u-aizu.ac.jp}
 \affiliation{Computer Solid State Physics Laboratory, University of Aizu,
Aizu-Wakamatsu 965-8580, Japan}
\author{M.~S.~Shur} 
 \affiliation{Department of Electrical, Computer and Systems  Engineering, Rensselaer Polytechnic Institute, Troy 12180, U.S.A}

\date{\today}

\begin{abstract}
We study plasma effects 
in  a micromachined high-electron
mobility transistor (HEMT)
with the microcantilever serving as the gate using
the 
developed a  model. 
The model accounts for 
mechanical motion of the microcantilever 
and spatio-temporal variations
(plasma effects)
of the two-dimensional electron gas(2DEG)
system in the transistor
channel.
The microcantilever mechanical motion is described
in the
point-mass approximation. The 
hydrodynamic electron transport model is used 
to describe distributed
electron plasma phenomena in the 2DEG system.
Using the developed model, we calculated the response function
characterizing the amplitude microcantilever oscillations
and the output electric signal
as functions of the signal frequency and the bias voltage
for the devices with different parameters.
We find the voltage dependences of the frequency of the mechanical
resonance and its damping. 
In particular, it is
demonstrated that the amplitudes of the mechanical oscillations
and output electric signal exhibit pronounced maxima
at the bias voltages close to the voltage of the 2DEG channel depletion
followed by a steep drop with further increase in the bias voltage.
\end{abstract}


\maketitle

\section{Introduction}
The concept of a field-effect transistor
with  a micromachined cantilever as floating gate
 was put forward and discussed a long time
ago by Nathanson {\it et al}~\cite{1} (see, also ref.~\cite{2}).
Recently~\cite{2}, this concept  was father evolved and 
a floating-gate high-electron-mobility transistor (HEMT) device
comprising a microcantilever over a two-dimensional 
electron gas (2DEG)
channel was fabricated and measured 
(see, for instance, also refs.~\cite{3,4,5,6,7}).
A theoretical assessment of this and similar devices
is usually based on equations governing the mechanical motion
 of the cantilever supplemented by some circuit equations.
However, the motion of the electrically actuated
 microcantilever results in a complex motion of electrons
in the 2DEG channel, which, in turn, leads to variations
of the electron sheet density and
the self-consistent  electric field affecting   the microcantilever
dynamics. The simplest effect of 2DEG in micromachined HEMTs is associated
with a finite conductivity of the latter, particularly at the voltages
corresponding to a significant depletion of the 2DEG channel.
The electron transient processes associated
with the variations of the electron sheet density and
the self-consistent  electric field,
i.e., electron plasma processes, can play a significant role
in HEMTs and different HEMT-based devices~\cite{8,9,10,11,12,13}
(see also experimental papers~\cite{14,15,16,17,18,19,20}). 
The plasma effects in the HEMT-like devices can result in 
resonant response at the frequencies coinciding with
the frequencies of plasma oscillations.
Since these frequencies usually fall into the terahertz range,
such effects are important in different terahertz devices. 
Despite a substantial difference in the resonant mechanical frequencies and
the characteristic plasma frequencies, 
micromachined elements such as  a microcantilever floating-gate may be
useful for terahertz devices. For instance, the microcantilever floating
gate can be used for the internal modulation of terahertz signals 
in the resonant detectors utilizing the excitation of plasma oscillations
(see, for example, ref.~ [8]).
The consideration of plasma effects in the HEMT-like devices requires 
a device model which adequately describes spatio-temporal variations of
the 2DEG system.

In this work, we develop a  model which self-consistently
describes
the mechanical oscillations of a highly conducting (metallized) 
microcantilever and the dynamic properties
of the 2DEG system. The model is based on an equation governing
oscillations of a microcantilever under mechanical and electric forces
and hydrodynamic equations governing the electron transport in 2DEG.
Our model accounts for such phenomena as the depletion and 
enrichment of 2DEG
by the applied voltage in the presence of
the surface charges at the semiconductor surface beneath the
microcantilever
and finiteness of the 2DEG conductivity,
 the delay in the electron recharging under the
microcantilever (gate)
and the 
spatial nonuniformity of the potential distribution in the 2DEG,
which might affect the device characteristics.
Thus, the model under consideration  is a distributed
(physical) model which provides more general and  detailed
description of the underlying processes than that based
on the treatment of the electron system of the device invoking electric
circuit models. 
Generally, the device under consideration is similar to those
fabricated and studied
both theoretically 
and experimentally, in particular, in refs.~\cite{1,2}.
The main distinction is that in micromachined HEMTs 
investigated in refs.~\cite{1,2} there are two metallized sufraces:
one at the bottom of the microcantilever and one covering
a portion of the semiconductor
surface under the microcantilever 
(the so-called input force plate~\cite{1}). In contrast, we assume
that the microcantilever motion is due to the interaction of
the charge induced in the
microcantilever metallized
surface and the charge in the 2DEG channel. Thus, the 2DEG
channel
plays the dual role: it is used for the microcantilever actuation
and the variation of
its conductivity is used to detect the output
signals. The effectiveness of the control of the
2DEG channel
by applying voltage to the microcantilever
was demonstrated experimentally
even in ref.~\cite{1}
(see, also ref.~\cite{2} and references therein).
The effect of interaction of the
charges induced in the microcantilever and the conducting plane electrode
was theoretically considered recently 
in ref.~\cite{21}.
However, the model used in ref.~\cite{21} assumes, 
in contrast to our model, that the electrode  
is ideally conducting that is not always the case in real 2DEG channels.

The developed model is used to study 
the resonant response of the device to the ac signals.
We find the dependences  of the resonance frequency and the resonance
 width 
on the both mechanical and electron properties of the system under 
consideration. In particular, we demonstrate that the plasma effects
in the 2DEG channel can give rise to a shift of the resonance
and an increase in
the oscillation damping, i.e.,  limit the 
quality
factor of the microcantilever oscillations.
We show also that relatively
high-frequency signals (with the frequency corresponding to the plasma resonance) can result in a significant variation 
of the microcantilever position.

The paper is organized as follows. In Sec.~2, we 
discuss the device model and write down the pertinent
equations. Section~3 deals with a small-signal analysis
based on the linearized versions of equations of the model.
In this section, we calculate the spatial distributions
of the ac potential and electron density in the 2DEG channel 
and find the amplitude (response function) of the microcantilever
oscillations as a function of the signal frequency
and the bias voltage. In Sec.~4, we analyze the response
function in different limiting cases
and demonstrate the results
of numerical calculations of
the device characteristics using the formulas
obtained in the previous sections.
Section~5 deals with the calculations
of the output source-to-drain ac current and the effective 
device transconductance. 
In Sec.~6, we consider the effect of
high-frequency signals ( on the variation 
of the microcantilever position associated with the plasma resonance.
In conclusion (Sec.~7), we draw the main results.
Some ancillary  calculations are factored out 
to Appendix A and Appendix B.

\begin{figure}
\begin{center}
\includegraphics[width=7.5cm]{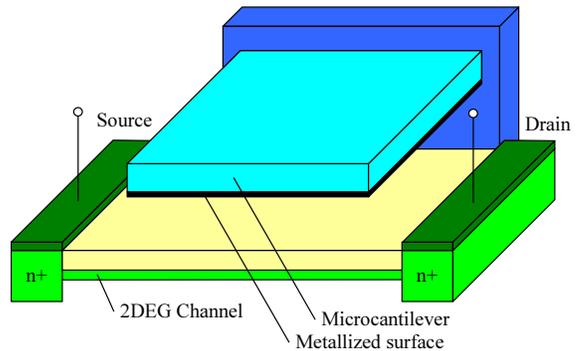}
\end{center}
\caption{ Schematic view of device  structure.
}
\end{figure}
\begin{figure}[t]
\begin{center}
\includegraphics[width=7.5cm]{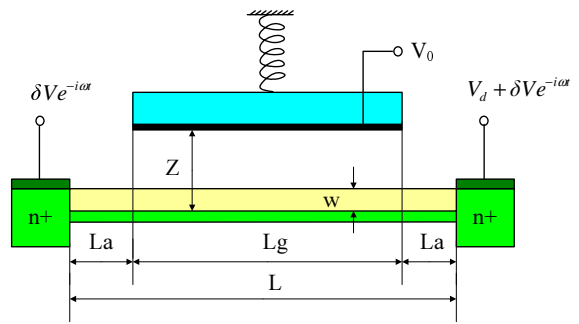}
\end{center}
\caption{Device model.
}
\end{figure}

\section{Equations of the model}

The device structure under consideration is schematically
shown in Fig.~1.
It is assumed that the voltage  applied between the metallized cantilever,
which serves as the HEMT floating gate, and the side contacts to 2DEG
channel (HEMT's source and drain) 
comprises the dc ($V_0$) and ac ($\delta V$)
components. 
Focusing on fairly detailed description of the electron
transport in the 2DEG channel accompanying the microcantilever oscillations,
we shall consider the cantilever mechanical properties in the framework
of a simplified model, namely, using the so-called 
point-mass model~\cite{22} (see Fig.2).
This model assumes that consideration
of  the elastic microcantilever or beam  is replaced by the consideration
of a point mass $M$ (cantilever effective mass)
attached to a string with stiffness $K$ with $M$ and $K$ chosen such that the resonant
frequency of the microcantilever
oscillations associated solely with its mechanical properties,
$\Omega_0 = \sqrt{K/M}$.
In this model,
the displacement of the cantilever (gate)
is governed by the following equation:
\begin{equation}\label{eq1}
M\biggl[\frac{\partial^2 Z}{\partial t^2} + \gamma_0\frac{\partial Z}{\partial t}
+ \Omega_0^2(Z - W)\biggr] = eD
\int_{-L_g/2}^{L_g/2} dx {\cal E}(\Sigma - \Sigma_d + \Sigma_s),
\end{equation}
where $Z = Z(t)$ is the distance between the cantilever surface
and the 2DEG, $W$ is this distance in the absence of
the applied voltage in equilibrium, $\gamma_0$ is the damping of
the cantilever oscillations associated with
different mechanisms of the energy loss in the cantilever body
and in the clamp, $e = |e|$ is the value of the electron charge,
$D$ and
$L_g$ are the pertinent sizes of the cantilever (see, Fig.2),
$\Sigma = \Sigma(t,x)$ is the electron sheet density of 2DEG,
 $\Sigma_d = const$ is the donor sheet density,
and 
$\Sigma_s = q_s/e$, where $q_s$ is the sheet density of 
the surface charge
at the interface between the semiconductor and the gas (or vacuum), which does not
change when the microcantilever moves.
The surface charge corresponds to the surface
potential 
$V_s = 4\pi ew\Sigma_s/\ae$.
The electric field ${\cal E} = {\cal E}(t,x)$ at the microcantilever
plane is determined by the potential
drop $V_0 - V_s - \varphi$, where  
$\varphi = \varphi(t,x)$ is the potential of  2DEG.
Here, the axis $z$ is directed perpendicular to the 2DEG plane,
while the axis $x$ is directed in the 2DEG plane.
Equation~(1) is valid in the case of a highly conducting (metallized)
cantilever when its surface is equipotential.
The microcantilever and 2DEG are separated
by two layers: the isolation solid layer of thickness
$w$ and dielectric constant 
$\ae$ and the layer of a gas (or vacuum) 
of thickness
$Z - w$ and dielectric constant 
$\ae^* \simeq 1$ (see Figs.~1 and 2).
The term in the right-hand side of eq.~(1) represents
the electric force acting on the microcantilever due the applied voltage.

In the gradual channel approximation valid if $Z, W \ll L_g$,~\cite{23}
\begin{equation}\label{eq2}
{\cal E} =  \frac{(\varphi - V_0 + V_s)}{[Z - w(1 - \ae^*/\ae)]}
 = - \frac{4\pi e}{\ae^*}(\Sigma - \Sigma_d + \Sigma_s).
\end{equation}
Equations~(1) and (2) disregard the effect of fringing capacitance,~\cite{23}
which can appear at the bias voltages beyond the essential depletion
of the gated portion of the 2DEG channel (this voltage range is not considered here).

Equations~(1) and (2) lead to
\begin{equation}\label{eq3}
\frac{\partial^2 Z}{\partial t^2} + \gamma_0\frac{\partial Z}{\partial t}
+ \Omega_0^2(Z - W) = - \frac{ D}{4\pi M}
\int_{-L_g/2}^{L_g/2} dx {\cal E }^2,
\end{equation}
where $Z \geq w$ and $\Sigma \geq 0$.
In the following, we put for simplicity $\ae^* = 1$.

Equations (2) and (3) should be supplemented by
the hydrodynamic equations (continuity equation and Euler equation)
governing the electron transport in 2DEG~\cite{8,9}:

\begin{equation}\label{eq4}
\frac{\partial \Sigma}{\partial t} + \frac{\partial\Sigma u }{\partial x} = 0,
\end{equation}
\begin{equation}\label{eq5}
\frac{\partial u}{\partial t} + u\frac{\partial u }{\partial x} + \nu u
= \frac{e}{m}\frac{\partial \varphi}{\partial x}.
\end{equation}
Here, $u = u(t,x)$ is the average (hydrodynamic) electron velocity
in the 2DEG plane, $\nu$ is the electron collision frequency, and $m$ is 
the electron effective mass. The electron collision frequency is expressed 
via the 2DEG mobility $\mu$ as $\nu = e/m\mu$.

\section{Small-signal analysis}

Let us assume that the net voltage between
the cantilever (gate) and the side contacts to the 2DEG (source and drain)
apart from dc components ($V_0$ and $V_0 + V_D$) 
comprises also the ac component
$\delta V\exp(-i\omega t)$:
where 
$\delta V$ ($|\delta V| \ll V_0$) and $\omega$ are the amplitude
and frequency of the ac voltage. 
The drain-to-source voltage $V_D$
is assumed to be sufficiently small and corresponds to the linear
region of the HEMT operation.
Considering small oscillations of the cantilever and the electron density,
one can assume
$$
Z(t) = Z_0 + \delta Z\,\exp(- i \omega t), 
$$
$$
\Sigma(t,x) = \Sigma_0 + \delta\Sigma\,\exp( -  i\omega t), 
$$
$$
 u(t,x) = \delta u\,\exp( - i\omega t), 
$$
$$
\varphi(t,x) = \delta \varphi\,\exp( - i\omega t), 
$$
where
the amplitudes  $\delta Z$, $\delta\Sigma$, $\delta u$,
and $\delta \varphi$ are assumed to be small in comparison with
the steady-state separation, $Z_0$, between the microcantilever and
the 2DEG channel (see Appendix A). 
Here $\Sigma_0 = \Sigma_d - \Sigma_s  + [(V_0 - V_s)/4\pi eZ_0]
\simeq \Sigma_d - \Sigma_s  + (V_0 /4\pi eZ_0)$ 
is the dc electron density in the gated portion 
of the 2DEG channel.
In the above equation
(and in the following), 
we have omitted for brevity the term with $V_s$ 
because of its smallness 
(it is proportional to
small values $w$ and $\ae^{-1}$: $V_s \propto w\Sigma_s/\ae)$.
Nevertheless, the direct
contribution of the surface charges to the dc electron density is
taken into account.
Considering the smallness of the abovementioned
amplitudes
of variations, eqs.~(1) - (5) can be linearized.
As a result, neglecting $w(1 - \ae^{-1})$ in comparison with
$Z_0$, we arrive at 
$$
 (\Omega_0^2 - i\gamma\omega - \omega^2)\delta Z = 
\biggl(\frac{ V_0^2}{2\pi  Z_0^2} \frac{DL_g}{ M}\biggr)\frac{\delta Z}{Z_0}
$$
\begin{equation}\label{eq6}
+ \biggl(\frac{ V_0^2}{2\pi  Z_0^2} \frac{DL_g}{ M}\biggr)
\frac{1}{L_g}\int_{-L_g/2}^{L_g/2} dx \frac{\delta\varphi}{V_0},
\end{equation}

\begin{equation}\label{eq7}
 \delta\Sigma = - \frac{1}{4\pi e Z_0}
\biggl(\delta\varphi+ V_0\frac{\delta Z}{Z_0}\biggr), 
\end{equation}

\begin{equation}\label{eq8}
-i\omega \delta\Sigma + \Sigma_0 \frac{d \delta u }{d\,x} = 0,
\end{equation}

\begin{equation}\label{eq9}
(\nu - i\omega) \delta u = \frac{e}{m}\frac{d \delta\varphi}{d\, x}
\end{equation}
\begin{equation}\label{eq10}
\frac{d^2 \delta\varphi}{d\, x^2} + 
\frac{m\omega(\omega + i\nu)}{4\pi e^2\Sigma_0Z_0}\delta\varphi
= - \frac{m\omega(\omega + i\nu)V_0}{4\pi e^2\Sigma_0Z_0}
\frac{\delta Z}{Z_0}
\end{equation}
or
introducing the  characteristic plasma velocity $S$ as
\begin{equation}\label{eq11}
S = \sqrt{\frac{4\pi e^2\Sigma_0Z_0}{m}},
\end{equation}
we obtain the following equations
$$
 \biggl[\Omega_0^2 - \Omega_0^2\biggl(\frac{V_0}{\overline{V_0}}\biggr)^2 - i\gamma_0\omega - \omega^2\biggr]\,\delta Z
$$
\begin{equation}\label{eq12}
 = 
\Omega_0^2 \biggl(\frac{V_0}{\overline{V_0}}\biggr)^2
\frac{Z_0}{L_g}\int_{-L_g/2}^{L_g/2} dx \frac{\delta\varphi}{V_0},
\end{equation}
and

\begin{equation}\label{eq13}
\frac{d^2 \delta\varphi}{d\, x^2} + 
\frac{\omega(\omega + i\nu)}{S^2}\delta\varphi
= - \frac{\omega(\omega + i\nu)V_0}{S^2}
\frac{\delta Z}{Z_0}.
\end{equation}

The boundary conditions for eq.~(13) can be chosen
considering that the amplitude of the ac potential of the side contacts
is equal to $\delta V$ and taking into account the ac
potential drop across the access sections of the channel (see Fig.~2),
i.e., the regions
 between the gate edges and the side contacts.
Assuming that the electron collision frequencies in different
regions of the 2DEG channel are the same and neglecting
the deviation of the electron sheet density
in the access regions from the donor density $\Sigma_d$,
the boundary conditions can be presented as~\cite{24}

\begin{equation}\label{eq14}
\delta\,\varphi|_{x = \pm L_g/2} = \delta V 
- L_a\biggl(\frac{\Sigma_0}{\Sigma_d - \Sigma_s}\biggr)\,
\frac{d\,\varphi}{d\,x}\biggr|_{x = \pm L_g/2}.
\end{equation}
Consider first the device structure in which the length, $L_a$,
of the access (ungated)
sections,  
is sufficiently small in comparison
 with the gate length $L_g$ (i.e.,
$L_a \ll L_g\Sigma_d/\Sigma_0$).
In this case, one can neglect the ac potential drop
across the access regions of the 2DEG channel, i.e., the second term
in the right-hand side of eq.~(14)~\cite{11,24}
As a result, from eqs.~(13) and (14) we obtain
$$
\delta \varphi = \frac{V_0}{Z_0}
\biggl[\frac{\cos\biggl[\sqrt{\omega(\omega + i\nu)}\,x/S]}
{\cos[\sqrt{\omega(\omega + i\nu)}\,L_g/2S]}- 1 \biggr]\,\delta Z 
$$
\begin{equation}\label{eq15}
+ 
\frac{\cos[\sqrt{\omega(\omega + i\nu)}\,x/S]}
{\cos[\sqrt{\omega(\omega + i\nu)}\,L_g/2S]}\,\delta\,V.
\end{equation}
Substituting $\delta \varphi$ from eq.~(15) to eq.~(12), we arrive at
\begin{equation}\label{eq16}
\frac{\delta\,Z }{\delta\,V} = \frac{Z_0}{V_0}\,{\cal Z}_{\omega},
\end{equation}
where
\begin{equation}\label{eq17}
{\cal Z}_{\omega} = 
\frac{\Omega_0^2(V_0/\overline{V_0})^2
(\tan Q_{\omega}/Q_{\omega})}{[\Omega_0^2 - i\gamma_0\omega - \omega^2 - 
\Omega_0^2(V_0/\overline{V_0})^2
(\tan Q_{\omega}/Q_{\omega})]}.
\end{equation}
Here, we have introduced
\begin{equation}\label{eq18}
Q_{\omega} =
\frac{\pi}{2}\frac{\sqrt{\omega(\omega + i\nu)}}{\Omega_p},
\qquad \overline{V_0} = \sqrt{\frac{2\pi  \Omega_0^2MW^3}{L_gD}},
\end{equation}
where 
\begin{equation}\label{eq19}
\Omega_p = \frac{\pi S}{L_g} = 
\sqrt{\frac{4\pi^3e^2\Sigma_0Z_0}{ m L_g^2}} =\Omega_{p0}
\sqrt{\frac{\Sigma_0Z_0}{(\Sigma_d - \Sigma_s)W}}
\end{equation}
is the characteristic plasma frequency of the gated  2DEG channel
and
$\Omega_{p0} = \sqrt{4\pi^3e^2(\Sigma_d - \Sigma_s)W/ m L_g^2}$

One can see that $\Omega_p$ depends on the bias voltage
via the voltage dependence of $\Sigma_0$ and $Z_0$.
The $\Omega_p$ versus $V_0$ dependence is asymmetric;
The plasma frequency can be significantly decreased by negative bias
when the 2DEG channel becomes close to the depletion.

The finiteness of the conductivity of the access ungated regions
can also contribute to the damping of the microcantilever oscillations.
This can occur when the length of this regions $L_a$ is sufficiently
large. To include these regions into the model, 
we need to modify
boundary condition~(19) to take into account the potential drop
across them.
Generally, the access regions can pronouncedly affect
the plasma phenomena in HEMTs~\cite{11,14,23},
in particular, leading to a decrease 
in the characteristic plasma frequencies.

Preserving the second term in the right-hand side of eq.~(4)
associated with the contribution of the access region to the boundary
conditions,
we arrive at 
\begin{equation}\label{eq20}
{\cal Z}_{\omega} = 
\frac{\Omega_0^2(V_0/\overline{V_0})^2
(\tan Q_{\omega}/Q_{\omega}^*)}
{\{\Omega_0^2 - i\gamma_0\omega - \omega^2 - 
\Omega_0^2(V_0/\overline{V_0})^2
(\tan Q_{\omega}/Q_{\omega}^*)\}},
\end{equation}
where $Q_{\omega}^* = Q_{\omega}(1 - \alpha Q_{\omega}\tan Q_{\omega})$
 and $\alpha = (2L_a/L_g)[\Sigma_0/(\Sigma_d - \Sigma_s)]$ is the parameter
characterizing the role of the access regions.
At $\alpha = 0$, $Q_{\omega}^* = Q_{\omega}$ and eqs.~(17) and (20) coincide.

\section{Microcantilever forced oscillations (Analysis of limiting cases
and numerical calculations)}

\subsection{Highly conducting 2DEG channel}

In many practical situations,
the signal frequency $\omega$ is
in the the same  range as  the resonant frequency 
of the microcantilever
oscillations $\Omega_0$ and the conductivity of the 2DEG channel
is rather large. The latter corresponds to  $\Omega_p \gtrsim \nu$. 
Since usually  $\omega, \Omega_0 \ll \nu, \Omega_p$,
the quantity
$|Q_{\omega}| \ll 1$, and  eqs.~(17) and (20)
 can be simplified.
In such a case,
one obtains 
$\tan Q_{\omega}/Q_{\omega} \simeq 1 + Q^2_{\omega}/3 \simeq 1
+ i(\pi^2/12)(\omega\nu)/\Omega_p^2)$. Considering this, assuming
that $\alpha \ll 1$ and using eq.~(17),
the response function ${\cal Z}_{\omega}$ can be presented in
the standard form

\begin{equation}\label{eq21}
{\cal Z}_{\omega}
\simeq \frac{\Omega_0^2}{(\Omega_m^2 - i\gamma_m\omega - \omega^2 )}
\biggl(\frac{V_0}{\overline{V_0}}\biggr)^2.
\end{equation}
Here, however, the resonant frequency $\Omega$ and 
the quantity characterizing the damping of oscillations $\gamma$ depend
on the ``electron'' parameters:
\begin{equation}\label{eq22}
\Omega_m = \Omega_0\sqrt{1 - \biggl(\frac{V_0}{\overline{V_0}}\biggr)^2},
\end{equation}
$$
\gamma_m = \gamma_0 + \nu\biggl(\frac{\pi^2}{12}\biggr)
\biggl(\frac{\Omega_0}{\Omega_p}\biggr)^2
\biggl(\frac{V_0}{\overline{V_0}}\biggr)^2
$$
\begin{equation}\label{eq23}
\simeq \gamma_0 + \nu
\frac{\displaystyle\biggl(\frac{\pi^2}{12}\biggr)
\biggl(\frac{\Omega_0}{\Omega_{p0}}\biggr)^2\biggl(\frac{V_0}{\overline{V_0}}\biggr)^2}{\displaystyle\biggl[1 + \beta\biggl(\frac{V_0}{\overline{V_0}}\biggr) -  
\frac{1}{2}\biggl(\frac{V_0}{\overline{V_0}}\biggr)^2\biggr]},
\end{equation}
where 
$\beta = {\overline{V_0}}/|V_0^{(depl)}|$ and
$V_0^{(depl)}$ is the characteristic depletion voltage
or HEMT's threshold
voltage. Here we have considered the dependence
of the characteristic plasma frequency on $\Sigma_0$ and
the dependence of the latter on the bias voltage (see Appendix B, eq.~(B2)).
%
As seen from eq.~(21), the resonant frequency is equal to $\Omega$.
The modulus of the response function $|{\cal Z}_{\omega}|$ 
is given by
\begin{equation}\label{eq24}
|{\cal Z}_{\omega}| = \frac{\Omega_0^2}
{\sqrt{(\Omega_m^2 - \omega^2)^2 + \gamma_m^2\omega^2}}\biggl(\frac{V_0}
{\overline{V_0}}\biggr)^2,
\end{equation}
so that at the resonance
\begin{equation}\label{eq25}
{\rm max}\,|{\cal Z}_{\omega}| = 
\frac{\Omega_0^2}{\gamma_m\Omega_m}\biggl(\frac{V_0}{\overline{V_0}}\biggr)^2.
\end{equation}
The second term in the right-hand side of eq.~(23) 
proportional to $\nu/\Omega_p^2 \propto \nu/\Sigma_0$ 
(i.e., proportional to the resistance
of the gated region of the 2DEG channel)
determines the contribution to the
resonance width associated with
the dissipation processes in the gated  2DEG channel
due to the finiteness of its conductivity. 
Equation~(24) demonstrates a tendency for $\gamma_m$ to increase with
approaching to the 2DEG channel depletion. 
If the second term in the right-hand side of eq.~(23)
becomes dominant, it might limit the quality
factor of the microcantilever oscillations: $Q_m = \Omega_m/\gamma_m < Q_0 =
\Omega_0/\gamma_0$. 
For example, for $\Omega_0/2\pi = 100$~MHz,
$\Omega_p/2\pi = 50$~GHz,  $\nu = 4\times 10^{12}$~$s^{-1}$
(electron mobility $\mu = 8000$~cm$^2$/V\,s), and $V_0/\overline{V_0}
= 0.1$, that can correspond
to a HEMT with GaAs channel close to depletion
at room temperature, the quality factor is limited 
by the value max$\,Q_m < 5\times 10^3$.

\subsection{Low conductivity of the 2DEG channel
(low plasma frequency)}
At a 
strong depletion of 2DEG channel at negative bias voltages,
the conductivity of 2DEG channel
and the plasma frequency can  be relatively  low.
At low plasma frequency $\Omega_p$ when
$\omega, \Omega_0 \gg \Omega_p^2/\nu$, one obtains $Q_{\omega} \simeq 
\pi\sqrt{i\omega\nu}/2\Omega_p$ with 
$|Q_{\omega}| = \pi\sqrt{  \omega\nu}/2\Omega_p \gg 1$.
In such a situation, 
$\tan Q_{\omega}/Q_{\omega} \simeq [(1 + i)\sqrt{2}/\pi]
(\Omega_p/\sqrt{\nu\omega})$. Hence, 
\begin{equation}\label{eq26}
{\cal Z}_{\omega} \simeq 
\frac{\displaystyle\Omega_0^2(V_0/\overline{V_0})^2
[(1 + i)\sqrt{2}\Omega_p/\pi\sqrt{\nu\omega}]}
{\Omega_0^2 - i\gamma_\omega - \omega^2 - \Omega_0^2(V_0/\overline{V_0})^2
[(1 + i)\sqrt{2}\Omega_p/\pi\sqrt{\nu\omega}]
}.
\end{equation}
As follows from eq.~(26), 
when $\Omega_0 \gg \Omega_p^2/\nu$ at the resonance (compare
with the pertinent formula for
the case highly conducting 2DEG channel),
\begin{equation}\label{eq27}
{\rm max}|{\cal Z}_{\omega}| \simeq  
\frac{\Omega_0}{\gamma_0}\biggl(\frac{2}{\pi}\frac{\Omega_p}{\sqrt{\nu\Omega_0}}
\biggr)\biggl(\frac{V_0}{\overline{V_0}}\biggr)^2 
\ll
\frac{\Omega_0}{\gamma_0}
\biggl(\frac{V_0}{\overline{V_0}}\biggr)^2.
\end{equation}

Thus, in the case under consideration here,
even at the resonance,
${\rm max}|{\cal Z}_{\omega}| \lesssim 1$. If $\sqrt{\nu/\Sigma_0}$,
increases, i.e., the 2DEG channel conductivity decreases,
${\rm max}|{\cal Z}_{\omega}|$ markedly decreases as well.
Hence, when the bias voltage $V_0$ approaches to the depletion
voltage $V_0^{(depl)}$, the  resonant peak markedly
diminishes. 

\subsection{Mechanical response at  plasma resonance}

If the signal and plasma frequencies markedly exceed
the electron collision frequency and the resonant frequency of pure
 mechanical oscillations of the microcantilever ($\omega, \Omega_p 
\gg \nu, \Omega_0$)
and
the signal frequency is close to one of the plasma frequencies
$\Omega_p(2n - 1)$, where $n = 1, 2, 3,...$ is the index of the plasma mode,
the quantity $|\tan Q_{\omega}/ Q_{\omega}|$ can be rather large.
At the fundamental plasma resonance
$\tan Q_{\omega}/ Q_{\omega} \simeq i(4/\pi^2)(\Omega_p/\nu)$.
Taking this into account,
from Eq.~(18) we find that as in the case considered 
in the previous subsection,
${\rm max}|{\cal Z}_{\omega}| \lesssim 1$.
One can also find
that when $\Omega_p/\nu$ increases to infinity, 
${\rm max}|{\cal Z}_{\omega}|$
formally tends to unity.
However, in most realistic cases ($\Omega_0 \lesssim \nu \ll \Omega_p$ and 
$V_0 < \overline{V_0}$),
\begin{equation}\label{eq28}
{\rm max}|{\cal Z}_{\omega}| \simeq\frac{4}{\pi^2}
\biggl(\frac{\Omega_0^2}{\nu\Omega_p}\biggr)
\biggl(\frac{V_0}{\overline{V_0}}\biggr)^2 \ll 1.
\end{equation}
This shows that even at rather sharp plasma resonance,
the amplitude of the microcantilever oscillations remains small
in comparison with the amplitude at the mechanical resonance
at $\omega = \Omega$. Indeed, the ratio of ${\rm max}|{\cal Z}_{\omega}|$
at $\omega = \Omega_p$  and at $\omega = \Omega$  can be estimated as
\begin{equation}\label{eq29}
\frac{{\rm max}|{\cal Z}_{\omega}|_{\omega = \Omega_p}}{{\rm max}|{\cal Z}_{\omega}|_{\omega = \Omega}} \simeq \frac{\pi^2}{4}
\frac{\gamma_0\Omega_0}{\nu\Omega_p} \ll 1.
\end{equation}
At the plasma resonance,  the amplitude of the ac electric field
in the gated region can be rather large. 

\subsection{Role of the access regions}

Consider now the case when $\alpha$ is not small invoking Eq.~(20).
When
$\omega, \Omega_0 \ll \nu, \Omega_p$, one obtains 
$|Q_{\omega}| = \pi\sqrt{ \omega\nu}/2\Omega_p \ll 1$ and
eq.~(20) can be simplified. 
In the vicinity of the resonance $\omega \simeq \Omega_m$,
where in the case uder consideration (compare with eq.~(22))
\begin{equation}\label{eq30}
\Omega_m = \Omega_0\sqrt{1 - \frac{(V_0/\overline{V_0})^2}{[1 +
\alpha^2(\pi/2)^4(\Omega_0\nu)^2/\Omega_p^4]}},
\end{equation}
one obtains
\begin{equation}\label{eq31}
{\cal Z}_{\omega} \simeq\frac{\Omega_0^2(V_0/\overline{V_0})^2}
{(\Omega^2 - i\gamma\omega - \omega^2)
[1  - i\alpha(\pi^2/4)(\omega\nu)/\Omega_p^2]}.
\end{equation}
Here 
$$
\gamma_m = \gamma_0 +  \nu\biggl[\frac{\pi^2(1 + 3\alpha)}{12}\biggr]
\frac{(\Omega_0/\Omega_p)^2}{[1 +
\alpha^2(\pi/2)^4(\Omega_0\nu)^2/\Omega_p^4]}
\biggl(\frac{V_0}{\overline{V_0}}\biggr)^2 
$$
\begin{equation}\label{eq32}
= \gamma_0 +  \nu\biggl[\frac{\pi^2(1 + 3\alpha)}{12}\biggr]
\frac{(\Omega_0/\Omega_p)^2}{(1 + \Omega_0^2\tau_a^2) }\biggl(\frac{V_0}
{\overline{V_0}}\biggr)^2
\end{equation}
is the quantity characterizing the damping of oscillations.
Here $\tau_a = \alpha (\pi^2/4)\nu/\Omega_p^2 $
is the RC-delay time:  $\tau_a = R_aC_g$, where 
$R_a = (m\nu L_a/2e^2\Sigma_d)$ and 
$C_g = L_g/4\pi Z_0$ are the resistance of the access regions
and the effective capacitance of the gated portion
of the 2DEG channel, respectively.
When $\alpha(\pi^2/4)(\Omega_0\nu)/\Omega_p^2 = \Omega_0\tau_a \sim 1$,
eq.~(32) results in 
\begin{equation}\label{eq33}
\gamma_m \simeq \gamma_0 + \nu\biggl(\frac{\pi^2\alpha}{8}\biggr)
\biggl(\frac{\Omega_0}{\Omega_p}\biggr)^2
\biggl(\frac{V_0}{\overline{V_0}}\biggr)^2
\end{equation}
One can see that the second term in the right-hand side of eq.~(33) differs
from the pertinent term in eq.~(23) by a factor $3\alpha/2$ which can be large
(when $L_a \gg L_g$).

Equation~(31) leads to
\begin{equation}\label{eq34}
|{\cal Z}_{\omega}| \simeq 
\frac{\Omega_0^2}
{\sqrt{(\Omega_m^2 - \omega^2)^2 +\gamma_m^2\omega^2   
}\,\sqrt{1 + \omega^2\tau_a^2}}\biggl(\frac{V_0}{\overline{V_0}}\biggr)^2.
\end{equation}
At the exact mechanical resonance $\omega = \Omega_m$,
from eq.~(34), we obtain
\begin{equation}\label{eq35}
{\rm max}|{\cal Z}_{\omega}| \simeq 
\frac{\Omega_0}{\gamma_m\sqrt{1 + \Omega^2\tau_a^2}}\biggl(\frac{V_0}{\overline{V_0}}\biggr)^2.
\end{equation}

One needs to point out that  the second terms in the right-hand side
of  eqs.~(24) and (34)  are the  products of  small factors,
$(\Omega_0/\Omega_p)^2$ and 
$(V_0/\overline{V_0})2$,  and relatively
large value $\nu$ (normally $\nu \gg \gamma_0$). 

\subsection{Results of numerical calculations}

Figures~3  shows the
modulus of the response function $|{\cal Z}_{\omega}|$
determined by Eq.~(20) versus signal frequency calculated for  AlGaAs/GaAs
micromachined HEMT devices under consideration at different bias voltages.
The device parameters are chosen as follows:
$\Omega_0/2\pi = 100$~MHz, $Q_0 = \Omega_0/\gamma_0 = 2500$,
$\Omega_{p0}/2\pi = 1$~THz, and
$\nu = 10^{12}$~s$^{-1}$.
The above parameters 
correspond a AlGaAs/GaAs HEMT-based device with 
to $L_g = 2~\mu$m, $L_a = 1~\mu$m, $W = 0.5~\mu$m, 
$M/L_gD = 4\times 10^{-12}$~g/$\mu$m$^2$,
$\Sigma_d - \Sigma_s = 1\times10^{11}$~cm$^{-2}$, and 
the electron mobility
$\mu = 3\times 10^4$~cm$^2$/Vs. 
At the above parameters,
$\Omega_{p0}/\Omega_0 = 10^4$, $\nu/\Omega_0 = 10^4/2\pi$,
$\overline{V_0} = 33$~V, 
$V_0^{(depl)} = - 9$~V (so that $\beta = 3.67$), 
and $V_0^{(pull-in)} = 18$~V.
It is seen from Fig.~3 that the resonant frequency decreases
with increasing bias voltage (in line with eq.~(22)).
Figure~3 also shows that $|\cal{Z}_{\omega}|$ increases with
increasing $V_0$ reaching a maximum
at the voltage slightly smaller than the depletion
voltage $V_0^{depl}$. Further increase in $V_0$ leads to a drastic drop
in $|\cal{Z}_{\omega}|$.
This is confirmed by
Fig.~4  which shows the voltage dependence of max~$|{\cal Z}_{\omega}|$
calculated for the above parameters (curve 1).
The results of calculations for a device with 
 $\Sigma_d - \Sigma_s =3\times 10^{11}$~cm$^{-2}$
and $L_g = 2~\mu$m) are shown in Fig.~4 as well. 
The parameters related to curve 1  in Fig.~4
correspond to $|V_0^{(depl)}| < V_0^{(pull-in)}$, in contrast
to curve 2 for which  $|V_0^{(depl)}| > V_0^{(pull-in)}$.

\begin{figure}[t]
\begin{center}
\includegraphics[width=7.5cm]{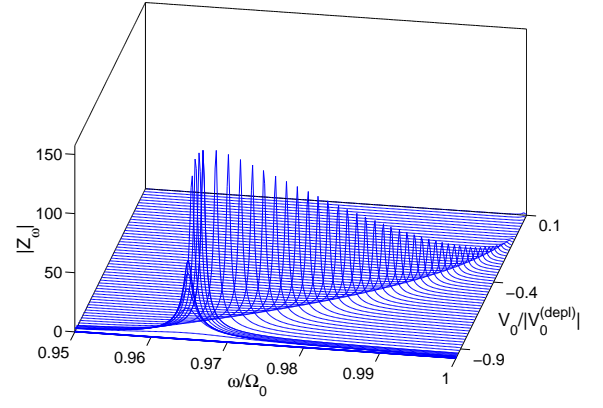}
\end{center}
\caption{ Modulus of the response function $|{\cal Z}_{\omega}|$
as a function of normalized
 signal frequency $\omega/\Omega_0$
 at different  bias voltages $\beta\, V_0/\overline{V_0} = V_0/|V_0^{(depl)}|$.  
}
\end{figure}
\begin{figure}[t]
\begin{center}
\includegraphics[width=7.5cm]{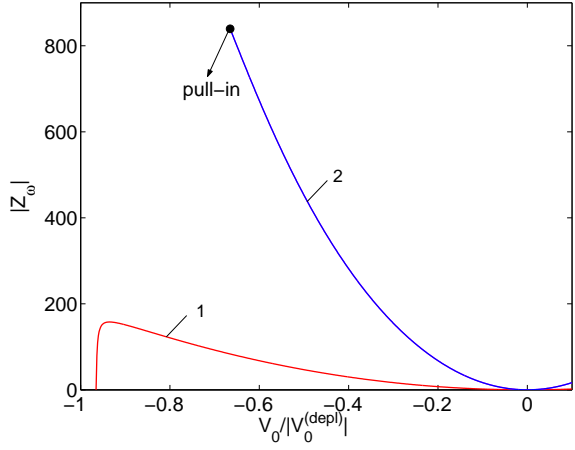}
\end{center}
\caption{Maximum (resonant) value of   $|{\cal Z}_{\omega}|$
as a function of normalized bias voltage 
$\beta\, V_0/\overline{V_0}= V_0/|V_0^{(depl)}|$
calculated for different structural parameters:
1 - $\Sigma_d - \Sigma_s =1\times 10^{11}$~cm$^{-2}$ and 
2 - $\Sigma_d - \Sigma_s =3\times 10^{11}$~cm$^{-2}$. 
}
\end{figure}

\section{Output electric signal}

The ac voltage applied between the microcantilever and the source
and drain contacts and the microcantilever oscillations
both result in the modulation of the electron density
in the 2DEG channel and, therefore, in the modulation
of the source-to-drain current.
Indeed, using eqs.~(7) and (15),
one can obtain (for the case of short access regions and relatively
small signal frequencies) 
$$
\delta \Sigma \simeq - 
\frac{(1 +{\cal Z}_{\omega}) }{4\pi eZ_0}\frac{\cos[\sqrt{\omega(\omega
+ i\nu)}\,x/S]}{\cos[\sqrt{\omega(\omega
+ i\nu)}\,L_g/2S]}\,\delta V 
$$
\begin{equation}\label{eq36}
\simeq
- 
\frac{(1 +{\cal Z}_{\omega}) }{4\pi eZ_0}
\frac{\cos[\sqrt{i\omega\nu}\,x/S]}{\cos[\sqrt{i\omega\nu}\,L_g/2S]}\,
\delta V \simeq - 
\frac{(1 +{\cal Z}_{\omega}) }{4\pi eZ_0}
\,\delta V.
\end{equation}
At the signal frequencies close to the microcantilever
resonant frequency, i.e., at the frequencies
significantly lower than the HEMT characteristic
frequencies, one can use the transistor steady-state characteristics.
As a result,
the ac component of the drain current, which can be considered as
the output signal, at low drain-to-source voltages $V_d  \ll |V_0|$
is given by
\begin{equation}\label{eq37}
\delta J_d = \frac{e\mu V_d}{L_g}\,\delta \Sigma.
\end{equation}
Using eqs.~(36) and (37), one obtains
\begin{equation}\label{eq38}
\delta J_d = - 
 \frac{g_0V_d}{4\pi e\Sigma_dZ_0}\,(1 + {\cal Z}_{\omega})\,\delta V
\simeq\frac{ g_0V_d}{V^{(depl)}}\,(1 + {\cal Z}_{\omega})\,\delta V, 
\end{equation}
where $g_0 = e\mu\Sigma_d/L_g$ 
is the conductance of the undepleted 2DEG channel.
Accounting for eq.~(17), from eq.~(38) we obtain
the following formula for the
effective transconductance 
$g_m = - (\partial J_d/\partial V)|_{V_d = const}$ of 
the micromachined HEMT under consideration:
$$
\frac{g_m}{g_0 } = \frac{V_d}{V^{(depl)}}\,(1 + {\cal Z}_{\omega})
$$
\begin{equation}\label{eq39}
= \frac{V_d}{|V^{(depl)}|}\,
\biggl[1 + \frac{\Omega_0^2(V_0/\overline{V_0})^2
(\tan Q_{\omega}/Q_{\omega})}{[\Omega_0^2 - i\gamma_0\omega - \omega^2 - 
\Omega_0^2(V_0/\overline{V_0})^2
(\tan Q_{\omega}/Q_{\omega})]}\biggr].
\end{equation}
The effective transconductance $g_m$ includes both the usual
component associated with the direct
electron density modulation by the ac voltage
 and the component associated with the microcantilever
oscillations (proportional to ${\cal Z}_{\omega}$).
In the limit of highly conducting 2DEG channel, eq.~(39) can be presented as
\begin{equation}\label{eq40}
\frac{g_m}{g_0 }
\simeq \frac{V_d}{|V^{(depl)}|}\,\biggl[1 + \frac{\Omega_0^2}{(\Omega^2 - i\gamma\omega - \omega^2 )}\biggl(\frac{V_0}{\overline{V_0}}\biggr)^2\biggr].
\end{equation}
As seen from eq.~(40), at the resonant frequency,
the modulus of the transconductance exhibits a 
rather high maximum:
\begin{equation}\label{eq41}
 {\rm max}\frac{|g_m|}{g_0 }\simeq \frac{V_D}{|V^{(depl)}|}
\frac{\Omega_0^2}{\gamma\Omega}\biggl(\frac{V_0}{\overline{V_0}}\biggr)^2
\simeq  \frac{V_D}{|V^{(depl)}|}
\frac{\Omega_0}{\gamma_0}\biggl(\frac{V_0}{\overline{V_0}}\biggr)^2.
\end{equation}
Since  in the devices with high quality factor of mechanical
resonance 
$|{\cal Z}_{\omega}|$ at the resonant frequency
can be much larger than unity,
the output signal in the micromachined HEMT can significantly
exceed the output signal in HEMTs with solely electrical modulation
However, as the bias voltage is approached to the depletion voltage,
the contribution of the microcantilever oscillations to the transconductance
vanishes and the latter steeply drops to zero.
As follows from eq.~(41) and
Figs.~3 and 4, the mechanical resonance provides very
sharp peaks  of $|{\cal Z}_{\omega}|$ and, hence, the 
transconductance modulus.

\section{Electric-field oscillations near the plasma resonance}

Using eqs.~(2) and (15), one can find the ac electric 
field $\delta {\cal E}$ as a function of $\delta V$ and 
the signal frequency $\omega$.
As shown the previous section, in the 
range of signal frequencies
where the ac electric field can exhibit the plasma resonances 
($\omega \simeq \Omega_p \gg \Omega_0, \nu$),
the ac displacement of the microcantilever is relatively small.
Hence, one can neglect the first term in the right-hand side of eq.~(16)
and, as follows from eq.~(2), put 
$\delta {\cal E} \simeq \delta\varphi/Z_1$. 
Here, $Z_1 = \langle Z \rangle$ is the dc position of the microcantilever
under
the dc electric field between the microcantilever
and 2DEG channel and the average effect 
of the  ac  electric field (compare with eq.~(A2)):
\begin{equation}\label{eq42}
Z_1 = W -   \frac{ DL_g}{4\pi\Omega^2 M}
\biggl[V_0^2 + \frac{1}{L_g}
\int_{-L_g/2}^{L_g/2} dx \langle{\cal E }^2\rangle\biggr],
\end{equation}
where the symbol $\langle ... \rangle$ means averaging
over fast oscillations with the frequency $\omega \gg \Omega_0$.
As a result, we arrive at the following equation: 
\begin{equation}\label{eq43}
\delta {\cal E} \simeq 
\frac{\cos[\sqrt{\omega(\omega + i\nu)}\,x/S]}
{\cos[\sqrt{\omega(\omega + i\nu)}\,L_g/2S]}\,\frac{\delta\,V}{ Z_1}.
\end{equation}

Substituting $\delta {\cal E}$ from eq.~(43) into eq.~(42), averaging
over high-frequency oscillations, we obtain
\begin{equation}\label{eq44}
Z_1 \simeq  Z_0  - 
W{\cal F}_{\omega}
\biggl(\frac{\delta V}{\overline{V_0}}\biggr)^2
\biggl(\frac{W}{Z_1}\biggr)^2.
\end{equation}
Here
\begin{equation}\label{eq45}
 {\cal F}_{\omega} = 
\frac{1}{4L_g}\int_{-L_g/2}^{L_g/2}\,d\,x\,
\biggl|\frac{\cos[\sqrt{\omega(\omega + i\nu)}\,x/S]}
{\cos[\sqrt{\omega(\omega + i\nu)}\,L_g/2S]}\biggr|^2
\end{equation}

At relatively low dc and ac voltages
when  $(W - Z_0)/W \ll 1$ and  $(W - Z_1)/W \ll 1$ (see eq.~(A4)), 
we obtain the following
formula:
\begin{equation}\label{eq46}
\frac{Z_1}{W} \simeq  1   - 
\frac{1}{2} \biggl(\frac{V_0}{\overline{V_0}}\biggr)^2  - 
{\cal F}_{\omega}
\biggl(\frac{\delta V}{\overline{V_0}}\biggr)^2.
\end{equation}
The second and third terms in the right-hand side of eq.~(46)  are
associated with the microcantilever displacement due to
the bias voltage and the ac signal, respectively.
Assuming $\Omega_p \gg \nu$, we obtain 
\begin{equation}\label{eq47}
{\cal F}_{\omega} \simeq 
\frac{[1 + (2\Omega_p/\pi\omega)
\sin(\pi\omega/2\Omega_p)\cos(\pi\omega/2\Omega_p)]}
{8[\cos^2(\pi\omega/2\Omega_p) + (\pi\nu/4\Omega_p)^2]}.
\end{equation}

As follows from eq.~(47), the electric response function ${\cal F}$
exhibits sharp resonances at $\omega = \Omega_p(2n - 1)$,
where $n = 1, 2, 3,...$ is the resonance index, 
if $\Omega_p \gg \nu$.
In the vicinity of  the fundamental resonant frequency, i.e., at 
$\omega \simeq \Omega_p$, 
eq.~(41) yields
\begin{equation}\label{eq48}
{\cal F}_{\omega} \simeq
\frac{1}{2\pi^2}\frac{\Omega_p^2}
{[(\omega - \Omega_p)^2 + \nu^2/4]}.
\end{equation}
At this resonance, one obtains
\begin{equation}\label{eq49}
{\rm max} {\cal F}_{\omega} \simeq \frac{2}{\pi^2}
\biggl(\frac{\Omega_p}{\nu}\biggr)^2 \gg 1.
\end{equation}
Equation~(49) implies that the position of the microcantilever
can be very sensitive to the incoming ac signals if their frequency
is close to one of the plasma resonant frequencies and the quality
factor of the plasma oscillations 
$Q_p \propto \Omega_p/\nu \gg 1$.
Hence the micromashined HEMT under consideration can serve
as a mechanical resonant detector of microwave and terahertz radiation.

One  may expect
that at sufficiently strong ac signals,
 the microcantilever can be
pulled-in to the surface of
the isolating layer.
Assuming, for simplicity that $V_0 = 0$,
we obtain the following condition of the microcantilever pull-in
under the effect of the ac voltage:

\begin{equation}\label{eq50}
\biggl(\frac{\delta V}{\overline{V_0}}\biggr)^2 
\geq \frac{8}{27{\cal F}_{\omega}}.
\end{equation}
Using the estimate for ${\rm max} {\cal F}_{\omega}$,
we obtain
\begin{equation}\label{eq51}
{\rm min}\,\biggl(\frac{\delta V}{\overline{V_0}}\biggr)^2 
\geq \frac{4\pi^2}{27}\biggl(\frac{\nu}{\Omega_p}\biggr)^2 \propto
\frac{1}{Q_p^2},
\end{equation}
so that the minimum ac pull-in voltage
can be estimated as
\begin{equation}\label{eq52}
{\rm min}\, \delta V^{(pull-in)} \simeq \biggl(\frac{\nu}{\Omega_p}\biggr)
\overline{V_0}.
\end{equation}
Equation~(54) shows that when the quality factor of 
the plasma oscillations is large, the microcantilever pull-in might
occur at
fairly modest ac signals.
However, one may assume that
the real situation is more complex because
the transition of the microcantilever  to the  position
corresponding to its pull-in to the isolating layer
should be accompanied by the channel depletion and significant
change in the resonant plasma frequency. Due to this, the dynamic
of the microcantilever pull-in out of the scope of this paper.

\section{Conclusions}

We developed a  model for a micromachined HEMT
with the microcantilever serving as the gate. 
The model is based on an equation
of mechanical motion of the microcantilever in the
point-mass approximation
accompanied by hydrodynamic equations describing distributed
electron plasma phenomena in the 2DEG channel.
Using this model, we calculated the response function
describing the amplitude microcantilever oscillations
and the output electric signal
as functions of the signal frequency and the bias voltage
for the devices with different parameters.
We found the voltage dependences of the frequency of the mechanical
resonance and its damping. 
It was
demonstrated that the amplitudes of the mechanical oscillations
and output electric signal exhibit pronounced maxima
at the voltages close to the voltage of the 2DEG channel depletion. 
However, further increase in the bias voltage results in a drastic
drop of the mechanical and electrical response.
We showed also that at the frequency corresponding to the plasma
resonance the  ac electric field between the microcantilever
and the 2DEG channel can be rather strong. This can result
in a significant variation of the microcantilever position 
by incoming high-frequency (terahertz) signals.

 This work was partially
 supported by the  Grant-in-Aid for Scientific Research (S)
from the Japan Society for Promotion of Science, Japan.
The work at RPI was partially supported by the Office of Naval
Research, USA.

\section*{Appendix A. Stationary states and pull-in and depletion voltages}
\setcounter{equation}{0}
\renewcommand{\theequation} {A\arabic{equation}}

When the voltage applied between the gate and side contacts
is constant, eqs.~(2) - (5) yield 
$u = u_0 = 0$,
$\varphi = \varphi_0 = 0$, and

\begin{equation}\label{A1}
\Sigma_0 = \Sigma_d - \Sigma_s + \frac{V_0 - V_s}{4\pi e[Z_0 - w(1 - \ae^{-1})] }
\end{equation}
 with $Z_0$ governed
by the following equation:
\begin{equation}\label{A2}
Z_0 = W - \frac{L_gD}{4\pi  \Omega_0^2M} 
\frac{ V_0^2}{[Z_0 - w(1 - \ae^{-1})]^2}.
\end{equation}
Introducing
$$
\overline{V_0} = \sqrt{2\pi  \Omega_0^2MW(W^2/L_gD)},
$$
eq.~(2) can be presented in the following form:
\begin{equation}\label{A3}
\frac{Z_0}{W} = 1 - \frac{1}{2} 
\biggl(\frac{ V_0}{\overline{V_0}}\biggr)^2
\biggl[\frac{W}{Z_0 - w(1 - \ae^{-1})}\biggr]^2, 
\end{equation} 
The states with $Z_0$ satisfying eqs.~(A2) and (A3) exist if
$|V_0| \leq V_0^{(pull-in)}$,
where 
$$
 V_0^{(pull-in)} = 
\sqrt{\frac{8}{27}
\biggl[1 - \frac{w}{W}\biggl(1 - \frac{1}{\ae}\biggr)\biggr]^3}\,\overline{V_0}
$$
\begin{equation}\label{A4} 
\simeq \sqrt{\frac{8}{27}}\,\overline{V_0}
\end{equation}
is the so-called pull-in voltage~\cite{1}.
One can find that when $|V_0| > V_0^{(pull-in)}$, 
eq.~(A3),  does not have roots. In this case,
the only existing (and stable) stationary state
corresponds to the attachment of the microcantilever
to the isolation solid layer, i.e., to $Z_0 = w$.
At $|V_0| = V_0^{(pull-in)}$, the microcantilever position
is $Z_0 = Z_0^{(pull-in)}$, where
\begin{equation}\label{A5} 
\frac{Z_0^{(pull-in)}}{W} = \frac{2}{3} + \frac{1}{3}
\frac{w}{W}\biggl(1 - \frac{1}{\ae}\biggr) \simeq \frac{2}{3}.
\end{equation}
When $|V_0| < V_0^{(pull-in)}$, eq.~(A3) has two solutions:
one with $Z_0 < Z_0^{(pull-in)}$ (unstable) and one with
$Z_0^{(pull-in)} < Z_0 < W$ (stable).
At $|V_0| \ll V_0^{(pull-in)}$, the position of the microcantilever
in the stable  state is given by
\begin{equation}\label{A6} 
\frac{Z_0}{W} \simeq 1 - \frac{1}{2} \biggl(\frac{V_0}{\overline{V_0}}\biggr)^2.
\end{equation}
There exists the stable state (not governed by eq.~(A3)) 
with $Z_0 = w$ as well .

As follows from eqs.~(A2) - (A3), $Z_0$ decreases with increasing $|V_0|$
disregarding the sign of the gate voltage $V_0$.
However, at $V_0 < 0$, the 2DEG channel can be fully depleted,
so that$\Sigma_0 = 0$.
The depletion voltage (or HEMT's threshold voltage) $V_0^{(depl)}$
corresponding to $\Sigma_0 = 0$,
as can be found from eqs.~(A1) and (A3), is given approximately by 
\begin{equation}\label{A7} 
V_0^{(depl)} \simeq - 4\pi e\Sigma_dW.
\end{equation}
This voltage corresponds to
\begin{equation}\label{A8} 
\frac{Z_0^{(depl)}}{W} = 
1 - \frac{1}{2}\biggl(\frac{V_0^{(depl)}}{\overline{V_0}}\biggr)^2.
\end{equation}
The second terms in the right-hand side of Eq.~(A8) 
is rather small.
Indeed, if  $\Sigma_d = (1 - 10)\times10^{11}$~cm$^{-2}$,
$L_gD = 25~\mu$m$^2$, $W = 0.5~\mu$m, $M = 10^{-10}$~g,
and $\Omega_0/2\pi = 1 - 10$~MHz, we obtain 
$\overline{V_0} \simeq 33 - 330$~V, $V_0^{(depl)} \simeq 9 - 90$~V,
and $V_0^{(pull-in)} \simeq 18 - 180$~V. Since $\overline{V_0}$
and, hence,  $V_0^{(pull-in)}$ strongly decrease with decreasing $W$,
their values can be markedly smaller than those obtained in the above
estimates.
The ratio of$|V_0^{(depl)}|$ to $V_0^{(pull-in)}$
can be presented as
\begin{equation}\label{A9} 
\frac{|V_0^{(depl)}|}{V_0^{(pull-in)}} 
\simeq \sqrt{\frac{27\pi e^2\Sigma_d^2L_gD}{\Omega_0^2MW}}
\end{equation}

When $V_0 < 0$ and $|V_0| > V_0^{(depl)}$, the charge densities
in the microcantilever and channel do not change with
varying $V_0$; they are equal to $\Sigma_d$.
In this case,  eq.~(A3) should be replaced by

\begin{equation}\label{A10}
\frac{Z_0}{W} = 1 - \frac{1}{2} 
\biggl(\frac{V_0^{(depl)} V_0}{\overline{V_0}^2}\biggr)
\biggl[\frac{W}{Z_0 - w(1 - \ae^{-1})}\biggr]. 
\end{equation} 
As a result, for the pull-in voltage under the condition
of the channel depletion we obtain
\begin{equation}\label{A11}
V_0^{pull-in} = \frac{1}{2}\frac{\overline{V_0}^2}{V_0^{(depl)}}
\biggl[1 - \frac{w}{W}\biggl(1 - \frac{1}{\ae}\biggr)\biggr]^2
\simeq \frac{1}{2}\frac{\overline{V_0}^2}{V_0^{(depl)}}
\end{equation}
and (compare with eq.~(A5))
\begin{equation}\label{A12}
\frac{Z_0^{(pull-in)}}{W} = 
\frac{1}{2}\biggl[1 + \frac{w}{W}\biggl(1 - \frac{1}{\ae}\biggr)\biggr] 
\simeq  \frac{1}{2}.
\end{equation}



\section*{Appendix B. Voltage and mechanical control of the plasma resonances}
\setcounter{equation}{0}
\renewcommand{\theequation} {B\arabic{equation}}

The expression for the characteristic plasma 
frequency $\Omega_p$ of the gated  2DEG channel 
given by eq.~(20) is somewhat different
from that obtained previously for the standard HEMTs.
This frequency in micromachined HEMTs exhibits different
voltage dependence. 
As follows from eq.~(20), 
the resonant plasma frequency $\Omega_p$ depends not only
on the electron density in the 2DEG channel but also on
the spacing, $Z_0$, between
the microcantilever (gate) and 2DEG.  
 This opens up 
the possibility of a mechanical control
of
the plasma frequency and, hence, different characteristics
of the pertinent terahertz devices.  

One can obtain the following dependence
of the characteristic plasma frequency on 
the microcantilever
displacement and the bias voltage.

\begin{equation}\label{eqB1}
\Omega_p = \Omega_{p0}\sqrt{\frac{\Sigma_0Z_0}{\Sigma_dW}}
\simeq \Omega_{p0}
\sqrt{\biggl(1 + \frac{V_0}{|V_0^{(depl)}|}\biggr)\frac{Z_0}{W}}.
\end{equation}
If $V_0$ is fixed, the dependence of the characteristic
plasma frequency on the microcantilever displacement 
is given by eq.~(41): $\Omega_p \propto \sqrt{Z_0}$.
Taking into account that the microcantilever displacement
depends on the bias voltage (see 
eq.~(A6) from Appendix A),
we find the following voltage dependence of the characteristic
plasma frequency:
\begin{equation}\label{eqB2}
\Omega_p \simeq 
\simeq
\Omega_{p0}
\sqrt{1 + \beta \biggl(\frac{V_0}{\overline{V_0}}\biggr)-  
\frac{1}{2} \biggl(\frac{V_0}{\overline{V_0}}\biggr)^2},
\end{equation}
where $\beta = {\overline{V_0}}/|V_0^{(depl)}|$.
It differs from the pertinent formula for the standard HEMTs
by the third term under the square root in the right-hand side of
eq.~(B2).




\end{document}